# Raman Fingerprints of Graphene Produced by Anodic Electrochemical Exfoliation

Vaiva Nagyte[1], Daniel J. Kelly[2], Alexandre Felten[3], Gennaro Picardi[1], YuYoung Shin[1], Adriana Alieva[1], Robyn E. Worsley[1], Khaled Parvez[1], Simone Dehm[4], Ralph Krupke[4,5], Sarah J. Haigh[2], Antonios Oikonomou[6,7], Andrew J. Pollard[8], Cinzia Casiraghi[*1]

[1] Department of Chemistry, University of Manchester, Manchester M13 9PL, UK.
[2] Department of Materials, University of Manchester, Manchester M13 9PL, UK.
[3] Synthesis, Irradiation and Analysis of Materials (SIAM), University of Namur, Belgium
[4] Institute of Nanotechnology, Karlsruhe Institute of Technology, Karlsruhe 76021, Germany
[5] Department of Materials and Earth Sciences, Technische Universität Darmstadt, Darmstadt 64287, Germany
[6] National Graphene Institute, University of Manchester, Manchester M13 9PL, UK.
[7] The Institute of Photonic Sciences, Castelldefels 08860, Spain
[8] National Physical Laboratory, Teddington, Middlesex, TW11 0LW, UK

*E-mail:cinzia.casiraghi@manchester.ac.uk

Electrochemical exfoliation is one of the most promising methods for scalable production of graphene. However, limited understanding of its Raman spectrum as well as lack of measurement standards for graphene strongly limit its industrial applications.

In this work we show a systematic study of the Raman spectrum of electrochemically exfoliated graphene, produced using different electrolytes and different types of solvents in varying amounts. We demonstrate that no information on the thickness can be extracted from the shape of the 2D peak as this type of graphene is defective. Furthermore, the number of defects and the uniformity of the samples strongly depend on the experimental conditions, including post-processing. Under specific conditions, formation of short conductive trans-polyacetylene chains has been observed.

Our Raman analysis provides guidance for the community on how to get information on defects coming from electrolyte, temperature and other experimental conditions, by making Raman spectroscopy a powerful metrology tool.

Graphene, the first discovered 2-dimensional (2d) material, is attracting strong research interest because of its outstanding properties, which can be exploited in a wide range of applications.[1] One of the most used techniques for large scale production of graphene is based on solvent assisted exfoliation,[2–5] which allows preparation of 2d crystal dispersions. In particular, the most common techniques are liquid-phase exfoliation (LPE)[6] and electrochemical exfoliation (ECE).[7] Although both LPE and ECE offer a simple, low cost and scalable way to produce graphene, commercialization of graphene-based products is still limited due to the lack of metrology standards.[8–13] In order to overcome this problem, the community has introduced some guidelines to establish universal practices to allow a better classification of solution-processed 2d materials, based on specific parameters, such as thickness, size and surface chemistry.[10,11] However, characterization of such materials is currently very challenging because dispersions typically contain nanosheets with a wide distribution in size and thickness, in contrast to traditional colloidal dispersions. Moreover, such distributions strongly depend on the exfoliation method and parameters used.[9]

Raman spectroscopy is a simple, low cost and widely used technique for the characterization of graphene.[14–16] The Raman spectrum of graphene is composed of two main features, the G and the 2D peaks, which lie at ~1580 cm$^{-1}$ and ~2700 cm$^{-1}$, respectively.[17] The full width at half maximum of the 2D peak (FWHM(2D)) of 25 - 30 cm$^{-1}$ is typically used to identify a single-layer of graphene.[17] However, graphene produced by different methods can show very different Raman spectra: in the case of solution-processed graphene, the Raman analysis has to take into account other factors that can alter the Raman spectrum, such as the interaction with the solvent, the reduced flake size (e.g. effect of edges), defect formation, and the random re-stacking (i.e. loss of AB stacking of the starting graphite) of the flakes. These effects give rise to strong changes in the Raman spectrum, as compared to that measured on graphene produced by micro-mechanical exfoliation (MME).[17,18] For example, graphene produced by LPE typically has FWHM(2D) above 50 cm$^{-1}$ [6,19–23] in contrast to ~25 cm$^{-1}$ for MME graphene[17]. Furthermore, the D peak at ~1350 cm$^{-1}$, is also observed in graphene produced by LPE.[6,19–23]

In our group we have developed a simple protocol for qualitatively assessing the thickness distribution of graphene dispersions produced by LPE.[20–23] Nevertheless, the Raman characterization of ECE graphene has been rather limited: transmission electron microscopy (TEM) and atomic force microscopy (AFM) are typically used to analyze the graphene flake's morphology,[24] while the chemical composition and crystal structure are typically evaluated by X-ray photoelectron spectroscopy (XPS) and X-ray powder diffraction (XRD).[25,26] It was observed that there is a correlation between XPS analysis on oxygen-containing groups, layer spacing from XRD data and the D and G peak intensity ratio (I(D)/I(G)).[27,28] Thus, I(D)/I(G) has been proposed as a parameter to define the quality of ECE graphene and optimal ECE process conditions.[29] However, it is known that this ratio does not depend linearly on defect density,[30–32] and it changes with the excitation energy.[31] Thus, inaccurate conclusions could be derived by analyzing I(D)/I(G) only or by comparing I(D)/I(G) measured in different works using different laser excitations. Indeed, it was already observed that graphene samples having similar C/O ratios can have rather different I(D)/I(G) values in some cases.[33] Furthermore, XPS and XRD are bulk characterization techniques, while Raman spectroscopy can easily probe individual graphene sheets as the typical flake size is well above the laser spot size (~500 nm). Most of the research on characterization of ECE graphene[34–37] reports Raman spectra of a few selected flakes, which may not be representative of the dispersion – a statistical analysis is needed.[9] Moreover, the Raman spectrum of an ECE graphene flake is compared with that of graphite to show successful exfoliation of graphene,[38,39] but no further analysis of the Raman features is performed. Only Wang et al.[40] presented Raman spectroscopy results combined with optical contrast analysis to evaluate the statistical distribution of the graphene sheet thickness. However, the optical contrast may vary depending on the degree of functionalization of graphene and intercalation stage.

In this work we show a systematic Raman spectroscopy analysis of graphene produced by ECE under different conditions. We show that the Raman analysis developed for LPE graphene cannot be applied to graphene obtained by ECE. In particular, the thickness of graphene produced by anodic ECE cannot be extracted by Raman spectroscopy. We propose a detailed approach to characterize ECE graphene, which allows qualitative determination of the level of defects in ECE flakes and the uniformity of the material. Furthermore, we show that the properties of such graphene are strongly sensitive to *post-processing*, further highlighting the importance of characterizing the material during every step of production, which is the key to enable batch to batch reproducibility and industrial uptake.

RESULTS AND DISCUSSION

To evaluate the influence of the electrolyte and the solvent on ECE graphene properties, 12 samples were prepared under different conditions (Table 1 and Figure S2.1). Note that each sample was repeated at least twice to ensure reproducible results (Section S7.4). Graphite electrodes were first electrochemically exfoliated (See Methods in Section S1) using different electrolytes: sodium sulfate ($Na_2SO_4$ (aq.)) (0.5 M), potassium bisulfate ($KHSO_4$ (aq.)) (0.5 M), and ammonium sulfate (($NH_4)_2SO_4$ (aq.)) (0.5 M). Subsequently, the exfoliated material was sonicated in a given volume (50 ml or 100 ml) of either n-methyl-2-pyrrolidone (NMP) or a water ($H_2O$) and isopropyl alcohol (IPA) mixture (1:1 ratio, v/v) for 30 min. The resulting dispersion was centrifuged at 3500 rpm (903g) for 10 min to sediment the residual graphite. Then UV-Vis spectroscopy was used to get concentration and calculate the yield (Table 1, Section S3). The obtained supernatant was collected and the flakes were immediately drop-cast on silicon substrates covered with an oxide layer, then dried at 100 °C for 10 h in vacuum oven to help removal of remaining solvent. AFM and Raman characterization were then performed. Graphene produced by LPE and graphene oxide (GO) were also prepared (Methods) and characterized for comparison.

**Table 1. List of sample names and corresponding experimental conditions used during ECE.**

| Sample name | Electrolyte | Solvent | Volume (ml) | Concentration (mg/ml) |
| --- | --- | --- | --- | --- |
| $Na_2SO_4$+NMP+50 | $Na_2SO_4$ | NMP | 50 | 0.063 |
| $Na_2SO_4$+NMP+100 | $Na_2SO_4$ | NMP | 100 | 0.047 |
| $Na_2SO_4$+IPA+$H_2O$+50 | $Na_2SO_4$ | IPA/$H_2O$ | 50 | 0.074 |
| $Na_2SO_4$+IPA+$H_2O$+100 | $Na_2SO_4$ | IPA/$H_2O$ | 100 | 0.035 |
| $KHSO_4$+NMP+50 | $KHSO_4$ | NMP | 50 | 0.05 |
| $KHSO_4$+NMP+100 | $KHSO_4$ | NMP | 100 | 0.027 |
| $KHSO_4$+IPA+$H_2O$+50 | $KHSO_4$ | IPA/$H_2O$ | 50 | 0.015 |
| $KHSO_4$+IPA+$H_2O$+100 | $KHSO_4$ | IPA/$H_2O$ | 100 | 0.012 |
| $(NH_4)_2SO_4$+NMP+50 | $(NH_4)_2SO_4$ | NMP | 50 | 0.091 |
| $(NH_4)_2SO_4$+NMP+100 | $(NH_4)_2SO_4$ | NMP | 100 | 0.033 |
| $(NH_4)_2SO_4$+IPA+$H_2O$+50 | $(NH_4)_2SO_4$ | IPA/$H_2O$ | 50 | 0.03 |
| $(NH_4)_2SO_4$+IPA+$H_2O$+100 | $(NH_4)_2SO_4$ | IPA/$H_2O$ | 100 | 0.011 |

Figure 1a shows three representative Raman spectra of the ECE graphene flakes showing the characteristic peaks of solution processed graphene: the G, D and D' peaks, located at ~1580 cm$^{-1}$, ~1350 cm$^{-1}$, and ~1620 cm$^{-1}$, respectively.[6] The D peak is activated by structural defects,[30,31,41] or by the edges of the nanosheets having sizes comparable to or smaller than the laser spot.[42] New Raman features were also observed in some flakes (see flake 3, Fig. 1a): the G peak shows a shoulder at ~1539 cm$^{-1}$ featuring a smaller intensity peak ~1139 cm$^{-1}$. Note that the three Raman spectra are associated with flakes from the same dispersion, hence they have been produced under exactly the same experimental conditions. A visual inspection of the Raman spectra in Fig. 1a indicates differences in the flakes properties, confirming the need for a statistical analysis. Note that the extra peaks at ~1539 cm$^{-1}$ and ~1139 cm$^{-1}$ are typically observed in samples prepared with the smaller solvent volume (50 ml), thus these samples will be discussed separately.

Figure 1b shows the corresponding AFM images of the same three flakes. No significant morphological difference between the flakes can be observed, despite the different Raman spectra. Further AFM analysis (Section S4) shows that the determination of the number of layers for this material can be challenging. The flakes have lateral sizes between 1 and 20 μm (Section S5). TEM and electron diffraction analysis show the presence of mostly single and bilayers (Figure S6.1).

Let us now apply the Raman protocol developed for LPE graphene[20–23] to our samples. This is based on the fitting of the 2D peak: firstly, the 2D band region is fitted with a Lorentzian function and then by using the residual fitting coefficient ($R^2$), which reflects the quality of the fit, the approximate thickness of the LPE flake is deduced (Section S1). The $R^2$ values are typically plotted as a function of I(D)/I(G), as the smaller flakes (i.e. with higher I(D)/I(G)) are typically the thinnest ($R^2$ closer to 1):[26,28] this is because during sonication the material gets thinner, but also smaller in size.[18,44] Figure 2a compares I(D)/I(G) as a function of $R^2$ for graphene produced by LPE and ECE. The graphene flakes made by ECE all have similar I(D)/I(G), between 1.5 and 2.5 (representative spectra in Figure S7.1.1), in strong contrast with the data from LPE (Figure 2a and Figure S7.2.1a). Clearly, the LPE characterization protocol fails for ECE graphene. This is attributed to the different processes involved in LPE and ECE: the first is an isotropic process, i.e. the ultrasound exfoliation, performed for many days, produces a decrease in both thickness and lateral size. Therefore, thinner flakes are likely to have a smaller lateral size, which is typically well below the laser spot size (~500 nm), leading to a relatively strong D peak in the Raman spectrum.[6,42] On the other hand, the ECE route is based on ions intercalation and the production of gas, which allows separation of the layers in just a few minutes. After ECE, only a short sonication step (30 minutes) is needed to re-disperse the graphene flakes following removal of the electrolyte, so the size of the flakes is not affected by sonication, allowing for large (few micrometers in lateral size) flakes to be obtained, in contrast to LPE, whose flakes have size below 700nm.[44,45] Figure 2a indi-cates that the D peak in graphene produced by ECE is not activated by the edges, but by defects in the basal plane, attributed to the oxidative expansion of the graphite edges during ECE. [24,46,47]

As graphene produced by ECE is defective, a different Raman analysis needs to be used for its characterization. The Raman spectrum of defective graphene shows a 2-stage evolution pathway.[30] In Stage 1, starting from pristine graphene, the Raman spectrum evolves as follows: the D peak appears and I(D)/I(G) increases; the D' appears; all the peaks broaden, thus G and D' begin to overlap. In this stage, I(D)/I(G) can be used to estimate the amount of defects,[30,41] while I(D)/I(D') can be used to distinguish between different type of defects.[41] At the end of Stage 1, I(D)/I(G) starts decreasing due to the decay of D peak intensity as a result of the reduced number of ordered rings, whereas the intensity of G peak is unaffected as G peak only relates to $sp^2$ pairs.[48] As the number of defects keeps increasing, the Raman spectrum enters Stage 2, showing a marked decrease in the G peak position and increased broadening of the peaks; I(D)/I(G) sharply decreases to zero and second-order peaks are no longer well defined. Thus, I(D)/I(G) needs to be coupled with another Raman fit parameter, such as the FWHM of the D, G or 2D peaks,[31] as these Raman parameters show the highest sensitivity to all types of defects[41] and increase with increasing amount of defects.[30,31,41] Figure 2b shows A(D)/A(G), where A is the area under the peak representing the intensity, as a function of the FWHM. Note that the area and FWHM(G) have been selected to allow comparison with previous results.[31] This figure shows that all data points related to graphene produced by ECE belong to Stage 2. In addition to this, the FWHM(G) spans from 30 to 70 $cm^{-1}$, indicating that the quality of graphene can change dramatically (Figure S7.2.1b). However, the FWHM(G) never reaches the values associated to GO – hence, graphene produced by ECE is less defective than GO, in agreement with previous works.[49]

**Effect of the electrolyte (100 ml volume)**

We start by comparing the effect of the solvent and electrolyte for a fixed volume (100 ml). Figure 3 shows the FWHM(2D) distribution for all samples (Table 1): each dot in this plot represents the FWHM(2D) value from the individual ECE graphene flake Raman spectrum, the vertical line indicates the distance between the maximum and minimum values while the horizontal line and the box are illustrating the mean value and standard deviation (SD) respectively. We decided to focus on FWHM(2D) – note that as all peaks broaden with increasing level of defects, the FWHM(D) can also be used (Figure S7.1.2 and Figure S7.3.2). However, it is not recommended to use the FWHM(G) due to possible artefacts caused by the presence of the D' peak. Figure 3 and Table S7.3.1 shows that samples produced using $KHSO_4$ (aq.) stand out as they provide the narrowest FWHM(2D) distributions as well as the lowest mean values. This indicates that graphene produced by ECE with $KHSO_4$ (aq.) is the less defective (for fixed exfoliation conditions), and also the most uniform sample in terms of defect distribution. This could be attributed to the dissociation path of $KHSO_4$ salts in water: in the anodic ECE process, the anions intercalate into the graphite electrode and are the reactive species directly producing $SO_2$ and $O_2$ gases. In the case of $(NH_4)_2SO_4$ (aq.) and $Na_2SO_4$ (aq.), the dissociation is not so effective and they give rise to a broader range of active ions: $Na_2SO_4$ (aq.) dissociates into $HSO_4^-$, $SO_4^{2-}$, $NaSO_4^-$, and $Na^+$ and $(NH_4)_2SO_4$ (aq.) into $HSO_4^-$, $SO_4^{2-}$, $NH_4SO_4^-$, and $(NH_4)^+$, in contrast to $KHSO_4$, which dissociates only into $K^+$, $H^+$, $HSO_4^-$ and $SO_4^{2-}$. Based on our results, $NaSO_4^-$ and $NH_4SO_4^-$ species are not only responsible for the larger variation in the FWHM(2D), but also for the higher level of defects observed in many flakes, having FWHM(2D) much larger than 100 cm$^{-1}$. This is in agreement with a previous study,[25] where nanoscopic holes were observed when using $(NH_4)_2SO_4$ as electrolyte.

Furthermore, we observed that $(NH_4)_2SO_4$ (aq.) and $Na_2SO_4$ (aq.) give rise to salt formation on the surface of graphene (Figure 4a and c, Figure S6.2), even after copious washing with water, in agreement with Li et al.[50]. This was also confirmed with XPS where sodium was detected in the samples $Na_2SO_4$+NMP+100 and $Na_2SO_4$+IPA+$H_2O$+100 (Section S8). Note that we did not observe any salt formation with $KHSO_4$ (aq.) (Figure 4b), thus this electrolyte provides the best quality (smaller amount of defects and highest purity) and most uniform distribution of graphene in solution (under our experimental conditions).

**Effect of the solvent (100 ml volume)**

Figure 3 shows that the Raman spectra of graphene dispersed in NMP tend to have FWHM(2D) lower than those of samples dispersed in the mixed solvent, when produced with the same electrolyte. However, the difference between the mean values is not very significant (Table S7.3.1). On the other hand, the distribution in FWHM seems to be consistently larger for graphene dispersions in mixed solvent (average SD = 14.4 $cm^{-1}$) than in NMP (average SD = 11.0 $cm^{-1}$). This increase in the defect distribution may also explain why most of the graphene dispersions prepared by ECE are shown to be stable in IPA and water (1:1) (Figure S2.2), offering a non-toxic alternative to the use of NMP. The difference in defect density and uniformity observed with the change of solvent could be assigned to the presence of water during the short sonication step: the radicals formed during the ECE process could react with water and IPA giving rise to further oxidation of graphene.[51]

In order to further investigate the effect of the solvent, XPS was also used to analyze the chemical composition of the samples (Section S8). While a good agreement between Raman spectroscopy and XPS was observed for some samples, discrepancies were found in other samples, indicating that there is no clear correlation between the FWHM(2D) and C/O ratio (Figure S8.2). This may indicate that not all defects are associated with carbon-bound oxygen groups. [VN1]

**Effect of the solvent volume**

It was observed that *post-processing* after ECE in a reduced volume gives rise to new features in the Raman spectrum of graphene produced with $KHSO_4$ (aq.) and $(NH_4)_2SO_4$ (aq.) electrolytes: 38% of the flakes measured from those samples show two peaks at ~1539 cm$^{-1}$ and ~1139 cm$^{-1}$ in their Raman spectra (Fig. 1a, Table S7.3.1). We assign the peaks to the $\nu_1$ mode and $\nu_3$ mode of trans-polyacetylene (t-PA),[52] as confirmed by performing multi-wavelength Raman spectroscopy (Figure 5a): the $\nu_1$ and $\nu_3$ modes show a characteristic dispersion[53] with the excitation wavelength, caused by the presence of conjugated π bonding segments of different length and therefore different band gaps. The 1139 cm$^{-1}$ peak (Figure 5b and Figure S7.5.2a) shows a dispersion of ~20 cm$^{-1}$/eV, which is in close agreement with the literature value of 25 cm$^{-1}$/eV.[54] The second peak at ~1539 cm$^{-1}$ is red-shifted with increasing wavelength (Figure 5a), in agreement with previous results.[55–57] We note that the t-PA modes are not visible when measuring at 633 nm laser excitation, which indicates the presence of relatively short C=C conjugation length.[58,59] This is confirmed by the $\nu_3$ peak position, which is ~1139 cm$^{-1}$ at 514.5 nm. Based on theoretical calculations,[60] this position should correspond to carbon chains of less than 10 atoms.

In order to understand if t-PA is localized everywhere or only in particular regions of the flakes, we have performed dielectrophoresis (Section S1) to deposit the flakes between reference points, allowing us to perform AFM, Raman mapping and Scanning Electron Microscopy (SEM) *on the same flake*. Figure 5c shows an SEM image of a flake with t-PA features in its Raman spectrum and its corresponding Raman maps (Figure 5d and e) taken with a spatial resolution of 400 nm. Despite the limited resolution given by the far-field Raman measurements, it is clear that the t-PA signal is not uniform, i.e. it is localized only in a few regions of the flake. This area seems to match with the wrinkles observed in the SEM image in Figure 5c. Figure 5e shows the map of the FWHM(D) on the same flake: we can observe broadening of the D peak corresponding to the same areas where the t-PA signal is stronger, indicating that t-PA is formed in highly defective regions of the sample. Note that the Raman spectrum of t-PA is resonant: as such, the signal intensity depends on the conjugated chain length.[60] Thus, we are only able to probe chains with a fixed conjugated length by mapping at a fixed wavelength.

The Raman features of t-PA have been observed in several carbon nanomaterials.[55,57,61,62] However, the process behind t-PA formation has never been fully explained. Based on our results (see detailed discussion in Section S7.5), the t-PA formation must be related to the *post-processing* step, in particular to the short sonication used to disperse the graphene powder in the solvent. One of the major effects of sonication is the increase of local temperature: although the sonication step only takes 30 mins or so, this is typically carried out in standard bath sonicators with no temperature control. Furthermore, a local increase of the temperature may affect the small volume samples more than the large ones (starting from the same powder amount). This was confirmed by the drop in the number of flakes with t-PA signatures in their Raman spectra when using a sonicator connected to a chiller (Table S7.3.1). Therefore, we can conclude that the t-PA formation is driven by the heating produced during the sonication. Additional experiments were conducted further confirming the high sensitivity of the material to temperature (Section S7.5).

Our results show that *post-processing* after ECE is extremely important because this type of graphene is strongly sensitive to temperature. In particular, it is crucial to ensure a constant temperature, even for a short sonication, as this may lead to formation of short conductive chains. We remark that many works do not provide any detail on *post-processing* ECE steps (e.g. sonication conditions, volume of the solvent, etc.), which are likely to strongly affect the properties of the material produced.

CONCLUSIONS

In this work we have presented a systematic Raman analysis of electrochemically exfoliated graphene produced under different experimental conditions. We have demonstrated a simple Raman characterization protocol that allows evaluation and comparison of the quality of graphene dispersions, in terms of the amount of defects and sample uniformity. Our work shows that the properties of graphene produced by ECE are very sensitive to all processing steps: from intercalation to dispersion. Samples are very sensitive to heating, which increases the defect concentration, leading to the formation of t-PA.

Although Raman spectroscopy cannot be used to quantify the thickness of graphene produced by ECE, we have shown that this technique is still a fundamental tool in the development of metrology standards that can be applied to commercial dispersions of electrochemically exfoliated graphene.

## ASSOCIATED CONTENT

**Supporting Information**

This includes methods, photographs, yield calculation, AFM, TEM, Raman spectroscopy, XPS data analysis of samples and post-processing experiments. . This material is available free of charge via the Internet at http://pubs.acs.org.

## AUTHOR INFORMATION

**Corresponding Author**

*Email: (C.C.) cinzia.casiraghi@manchester.ac.uk

## ACKNOWLEDGMENT

This work is supported by the EPSRC in the framework of the CDT Graphene NOWNANO and the projects EP/K016946/1, EP/P009050/1 and EP/N010345/1. AJP, SJH, DJK, VN acknowledge funding from the National Measurement System of the Department of Business, Energy and Industrial Strategy, UK. RW acknowledges the Hewlett-Packard Company for financial support. AA , GP, SJH and DJK acknowledge support from the European Research Council (ERC) under the European Union's Horizon 2020 research and innovation program under grant agreement No 648417 and No 715502.

Figure 1. a) Representative Raman spectra measured on individual flakes produced by electrochemical exfoliation from the same sample (($NH_4$)$_2SO_4$+IPA+$H_2O$+50, see Table 1)), hence produced under exactly the same experimental conditions; b) their corresponding AFM image.

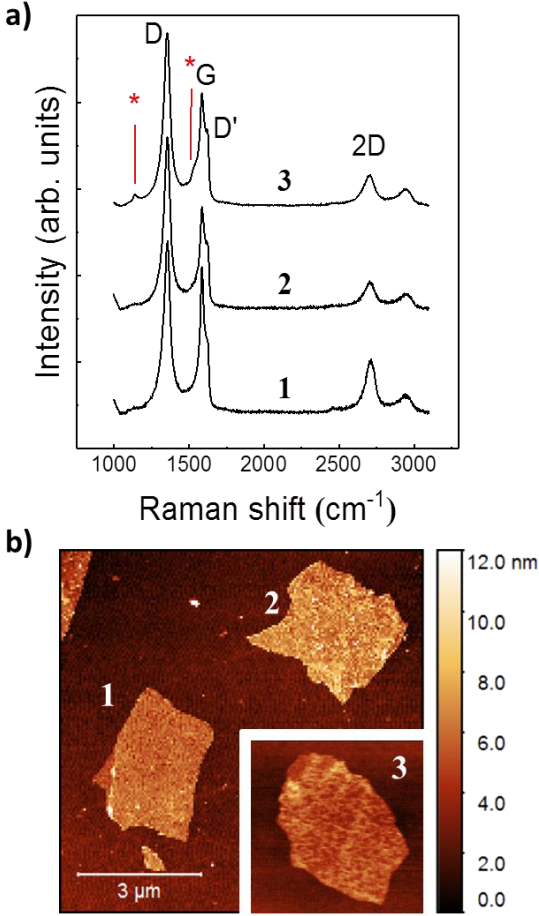

Figure 2. a) Raman analysis based on I(D)/I(G) and $R^2$ values of 2D peak fit, used for characterization of graphene produced by LPE, applied to ECE graphene (data collected from all samples). b) Two-stage defects trajectory based on A(D)/A(G) and FWHM(G), measured at 514 nm, for all graphene flakes produced by ECE, in comparison with graphene oxide (GO) and micro-mechanically exfoliated (MME) graphene.

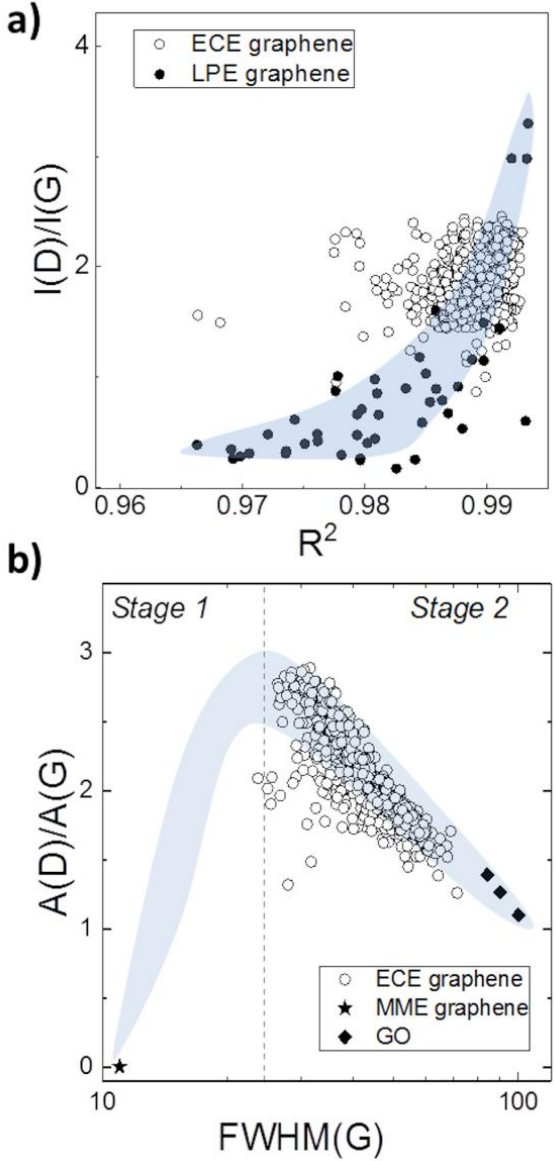

Figure 3. FWHM(2D) peak of the Raman spectra of graphene produced in 100 ml volume with different electrolytes and solvents (Table 1).

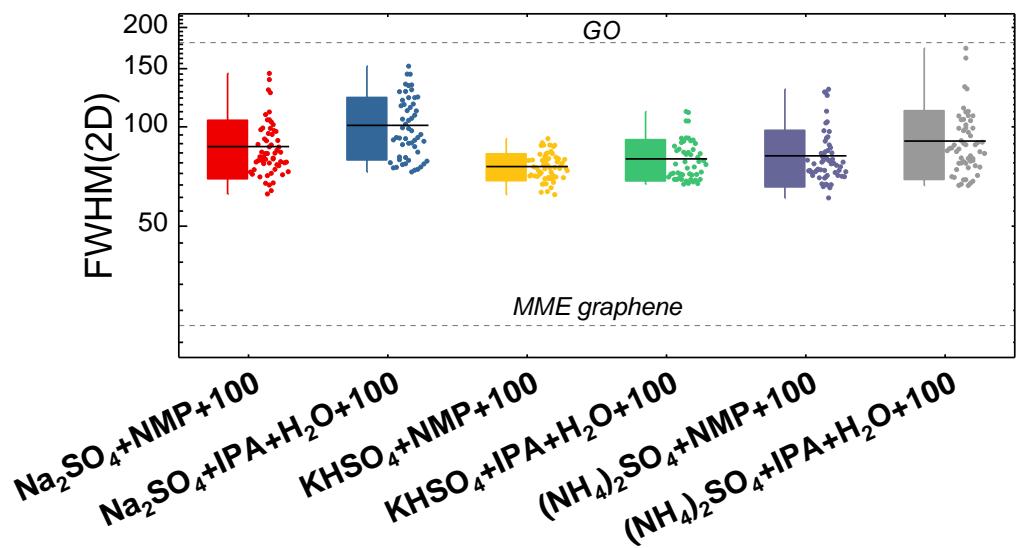

Figure 4. Scanning transmission electron microscopy images and EDS elemental mapping showing salt particles on flakes from samples a) $Na_2SO_4$+NMP+100 and c) $(NH_4)_2SO_4$+IPA+$H_2O$+50 while b) $KHSO_4$+NMP+100 did not show any presence of salts. Micrographs a) and b) are high-angle annular dark field (HAADF)-STEM, and c) is bright field-TEM. In (HAADF)-STEM salts are visible as spikey particles with higher brightness while in TEM they appear as regions of dark contrast.

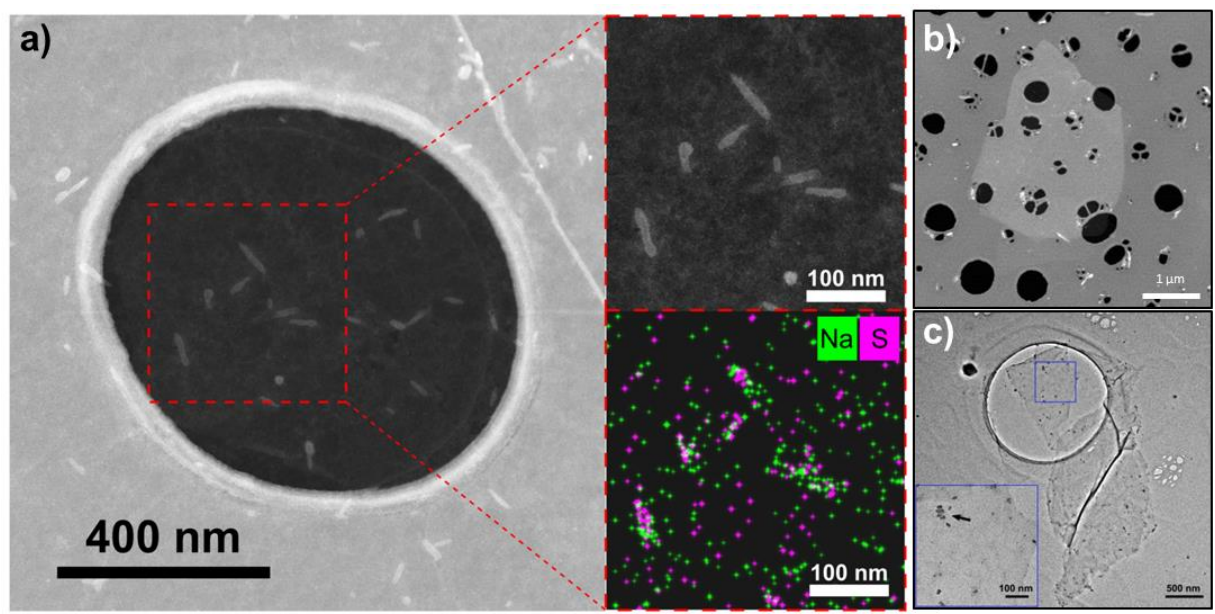

Figure 5. a) Raman spectra measured at different excitation wavelengths on a flake showing the t-PA peaks and b) related fits of t-PA of zoomed region in the spectra (a). c) SEM image of a flake (sample KHSO$_4$+NMP+50, Table 1) showing t-PA Raman peaks in its Raman spectrum; d-e) Raman maps measured on the flake in c) showing correlation between wrinkle presence on ECE graphene surface, d) intensity of 1539 cm$^{-1}$ peak and e) broadening of FWHM(D). These results show that t-PA is formed in region of high defectivity, which are likely to be associated with wrinkles.

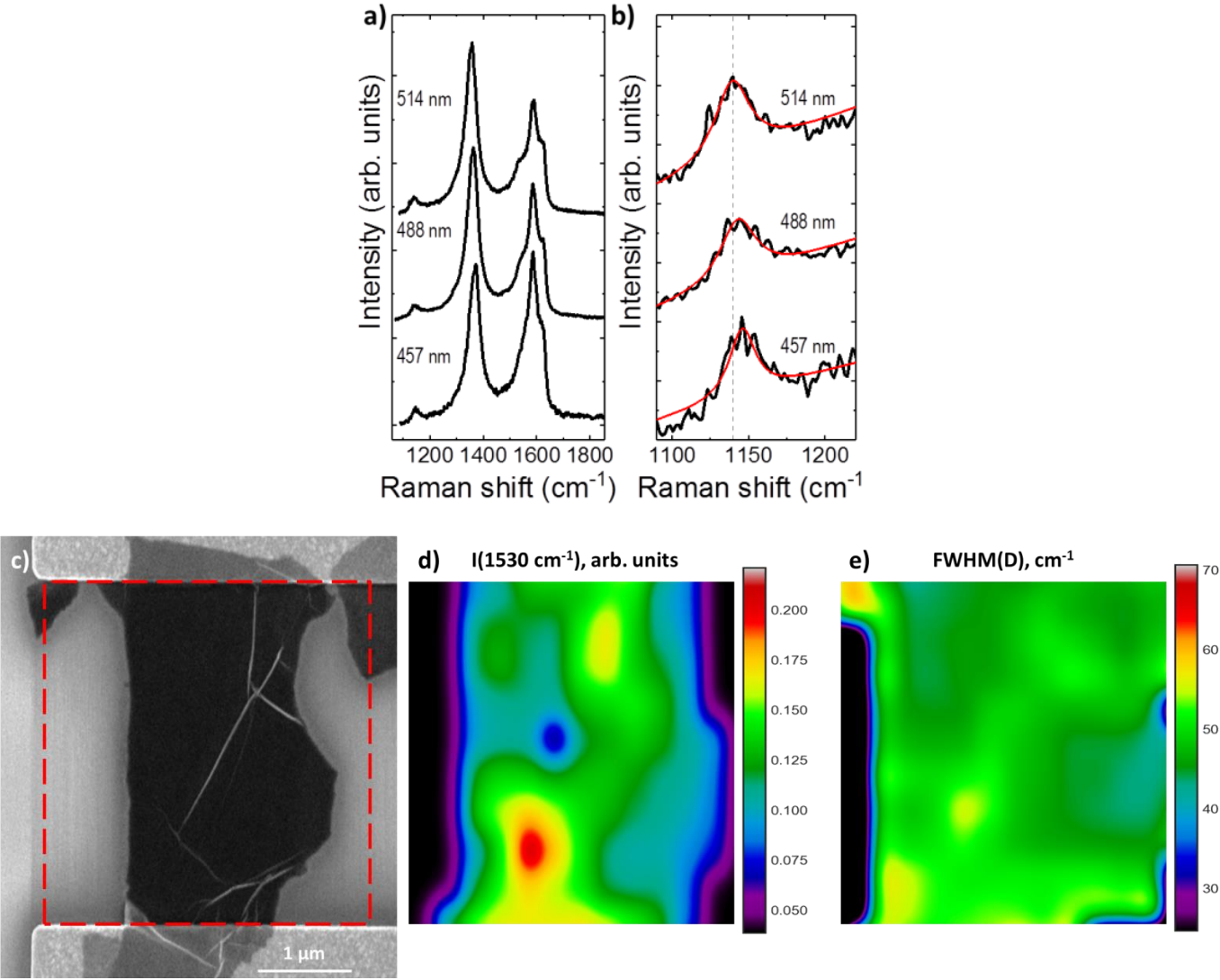

For Table of Contents Only

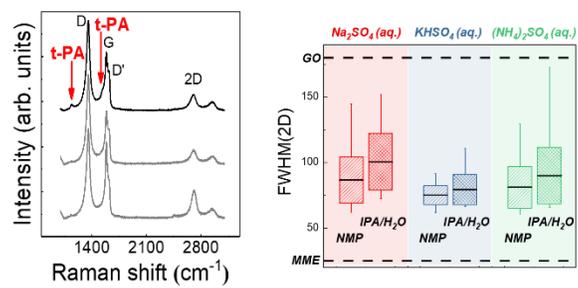